\documentclass[manuscript,screen,nonacm]{acmart}
\AtBeginDocument{%
  }

\usepackage{multirow} 
\usepackage{array} % required for text wrapping in tables
\usepackage{minted} % required for code snippets
\begin{document}

\title{From Correctness to Collaboration: Toward a Human-Centered Framework for Evaluating AI Agent Behavior in Software Engineering}

\author{Tao Dong}
\email{taodong@acm.org}
\author{Harini Sampath}
\authornotemark[1]
\email{harinis@google.com}
\author{Ja Young Lee}
\authornote{Equal contribution.}
\email{jayoung.j.lee@gmail.com}
\author{Sherry Y. Shi}
\authornotemark[1]
\email{sherryyshi@google.com}
\author{Andrew Macvean}
\email{amacvean@google.com}
\affiliation{%
  \institution{Google LLC}
  \country{USA}}


\begin{abstract}
As Large Language Models (LLMs) evolve from code generators into collaborative partners for software engineers, our methods for evaluation are lagging. Current benchmarks, focused on code correctness, fail to capture the nuanced, interactive behaviors essential for successful human-AI partnership. To bridge this evaluation gap, this paper makes two core contributions. First, we present a foundational taxonomy of desirable agent behaviors for enterprise software engineering, derived from an analysis of 91 sets of user-defined agent rules. This taxonomy defines four key expectations of agent behavior: \textit{Adhere to Standards and Processes}, \textit{Ensure Code Quality and Reliability}, \textit{Solving Problems Effectively}, and \textit{Collaborating with the User}.

Second, recognizing that these expectations are not static, we introduce the Context-Adaptive Behavior (CAB) Framework. This emerging framework reveals how behavioral expectations shift along two empirically-derived axes: the Time Horizon (from immediate needs to future ideals), established through interviews with 15 expert engineers, and the Type of Work (from enterprise production to rapid prototyping, for example), identified through a prompt analysis of a prototyping agent. Together, these contributions offer a human-centered foundation for designing and evaluating the next generation of AI agents, moving the field's focus from the correctness of generated code toward the dynamics of true collaborative intelligence.
\end{abstract}

\begin{CCSXML}
<ccs2012>
   <concept>
       <concept_id>10003120.10003121.10011748</concept_id>
       <concept_desc>Human-centered computing~Empirical studies in HCI</concept_desc>
       <concept_significance>500</concept_significance>
       </concept>
 </ccs2012>
\end{CCSXML}

\ccsdesc[500]{Human-centered computing~Empirical studies in HCI}

\keywords{Human-AI Collaboration, AI Agents, LLM Evaluation, Software Engineering}


\maketitle

\section{Introduction}
Powered by recent advancements in Large Language Models (LLMs), the nature of Generative AI tools for software engineering has undergone a fundamental transformation over the past few years, shifting from passive code completers \cite{vaithilingam2022expectation} to increasingly autonomous agents capable of tackling complex tasks including testing \cite{kang2023large}, migrations \cite{nikolov2025google}, bug fixing \cite{meng2024empirical, rondon2025evaluating} , and more. This evolution marks a critical paradigm shift: from AI as a tool to be wielded by a developer, to AI as a collaborative partner integrated into a software team's workflow \cite{kumar2025sharp, liu2024large}. As these agents start cooperating with human software engineers over multiple turns of actions, the success of the human-AI partnership hinges less on the raw correctness of any single generated code block and more on the quality of the collaborative problem solving process.

This rapid evolution from tool to partner has been heavily guided by the AI research community's primary mechanism for measuring progress: benchmarks. Serving as de facto goalposts, benchmarks define what "better" means and incentivize the development of more capable models. To this end, benchmarks like SWE-bench \cite{jimenez2024swebench}, LiveCodeBench \cite{jain2025livecodebench}, and Aider Polyglot \cite{AiderPolyglotBenchmark} have been undeniably successful. They have spurred a race to improve the functional correctness of AI-generated code, dramatically advancing LLMs' coding capabilities. 

Yet, this very success of benchmarks has created a critical blind spot. By fixating on the final output, these benchmarks are fundamentally unequipped to assess the behavioral qualities required for an effective partner. They ignore the crucial questions that arise in collaborative work: Does the agent adhere to project standards? Does it communicate its reasoning? Does it effectively incorporate user feedback? This disconnect from the reality of collaborative software development \cite{taylor2025software} has created a significant "evaluation gap."

Recognizing this evaluation gap, an emerging body of work has shifted focus from final outcomes to the agent's execution trajectory, analyzing the step-by-step process of its work \cite{jimenez2024swebench, deshpande2025trail, cemri2025multi}. While this is a crucial step forward, current trajectory analyses primarily focus on identifying mechanistic failures, such as incorrect tool usage or hallucinated commands. This approach, however, implicitly frames the agent's process through the lens of machine-centric errors rather than human-centric behaviors. It still treats the agent as a complex tool to be debugged, not a partner whose collaborative style must be assessed. This reveals a deeper, more fundamental problem: the research community lacks a systematic, user-grounded understanding of what constitutes desirable agent behavior in the first place. Before we can meaningfully evaluate the cooperative process, we must first define what a "good" process looks like from the software engineer's perspective.

To address this challenge and move beyond a monolithic view of agent performance, we ask the overarching research question: \textit{What are the core behaviors an intelligent agent should be evaluated for in software engineering? }Our work answers this by making two distinct but related contributions grounded in three research studies.

Our primary contribution is a foundational taxonomy of desirable behaviors for AI agents in contemporary enterprise software engineering. Derived directly from our analysis of 91 sets of real-world user-defined rules for customizing agent behavior (\hyperref[sec:study1]{Study 1}), this taxonomy provides a baseline of four core behavioral expectations that are critical for an agent's effectiveness today. Recognizing that any static taxonomy is inherently limited, our secondary contribution is an emerging framework for adapting expected agent behavior along two axes: \textit{Time Horizon} and \textit{Type of Work} (see Figure \ref{fig:cab_framework}). Its two analytical axes are empirically derived from our other studies. We establish its Time Horizon axis through semi-structured interviews with 15 expert engineers (\hyperref[sec:study2]{Study 2}), allowing us to map the critical headroom between current needs and the aspirational goals towards Artificial General Intelligence (AGI). We then define its Type of Work axis through an analysis of user prompts for an AI agent focused on rapid prototyping (\hyperref[sec:study3]{Study 3}), which reveals how behavioral expectations could shift when the task changes from production-grade engineering to UI-focused new product exploration. 

Together, our taxonomy and framework offer a nuanced, human-centered foundation for designing and evaluating the next generation of software engineering agents. They move the conversation beyond the monolithic metric of code correctness, providing the vocabulary and structure needed to assess true collaborative intelligence. In the remainder of this paper, we situate our work within prior research, detail the three studies that form the empirical basis of our contributions, and conclude by discussing how our framework can directly inform and enhance emerging trajectory-based evaluation methods—helping to finally close the gap between how we build AI partners and how we measure their success.

\section{Related Work}
Our work is situated at the confluence of three research areas: the benchmarking of Software Engineering (SWE) agents, the evaluation of human-AI collaboration, and the study of human expertise in software development. We argue that while each area provides a crucial piece of the puzzle, a significant gap exists at their intersection—a gap our work aims to fill.

\subsection{Evaluating SWE Agents}
The dominant paradigm for evaluating AI in software engineering has evolved alongside agent capabilities. Early benchmarks such as HumanEval \cite{chen2021evaluating} and MBPP \cite{austin2021program} established functional correctness as the primary metric, treating the agent as a black-box code generator. Success was measured by pass@k rates on unit tests, a suitable method for evaluating single-shot code completion. As tasks grew more complex, benchmarks like SWE-Bench \cite{jimenez2024swebench} and SWE-Lancer \cite{miserendino2025swelancer} maintained this focus on final outcomes, albeit for more challenging, multi-file problems.

Recognizing that functional correctness is only one facet of code quality \cite{usrey1996dimensions, takerngsaksiri2025code, green2024quality}, subsequent work introduced multi-dimensional evaluations. For example, Zheng et al. \cite{zheng2024beyond} proposed the RACE benchmark to assess code for Readability, mAintainability, Correctness, and Efficiency, reflecting a more holistic view of the final artifact.

Most recently, with the rise of autonomous agents in coding, the research focus has shifted from the final output to the process of generating it. Trajectory analysis has emerged as a key technique for understanding how an agent arrives at a solution. For example, SWE-Agent’s analysis of the agent’s failure modes helped its developers identify and make improvements to its tools and execution environment.

To guide failure mode analyses, Frameworks like MAST \cite{cemri2025multi} and TRAIL \cite{deshpande2025trail} provide valuable perspectives and key issues to look out for. MAST proposes a taxonomy of multi-agent system failures, focusing on agentic reasoning and inter-agent coordination, while TRAIL’s taxonomy covers failures in agentic reasoning as well as system-level failures like resource management and configuration errors. Bouzenia et. al’s empirical study on agent trajectories reveal some common patterns and anti-patterns of popular SWE agents that further our understanding of their process \cite{bouzenia2025understanding}. However, current trajectory analyses are fundamentally system-centric (e.g., tool calling or file editing failures), focusing on capabilities that are necessary but insufficient for the agent to be helpful to the user. 

\subsection{Evaluating AI’s Ability to Collaborate with Users}

One important capability of coding agents that many existing evaluation techniques do not assess is the agent’s ability to collaborate with the user. In contrast, in domains such as medicine and education, human-centric metrics have been adopted when assessing LLMs’ capabilities. For instance, studies in medicine evaluate AI not just for diagnostic accuracy but also for the quality and empathy of its communication with patients \cite{li2025counselbench, ayers2023comparing}. Similarly, in education, AI tutors are assessed on their ability to provide scaffolding and encouragement, not just correct answers \cite{maurya2024unifying, shi2025educationq, jurenka2024towards}.

Within the AI for SWE domain, researchers are beginning to acknowledge this need. In an observational study, Kumar et al. \cite{kumar2025sharp} found a direct correlation between successful task completion and the amount of human-agent communication, highlighting the importance of interaction quality. 

Evaluating interaction quality in an offline setting is a challenge, as there is no user for the agent to interact with. To address this challenge, benchmarks like MINT \cite{wang2023mint} and InterCode \cite{yang2023intercode} introduced a basic simulated user. Han et al. \cite{han2025convcodeworld} further evolved the idea of a simulated user by enabling varied responses and establishing personas with different expertise levels. While these approaches make interaction testable in offline settings, the personas themselves remain speculative. One way to ground these LLM-as-synthetic-users is to imbue them, through their system prompts, with the values and expectations of professional software developers, which our research seeks to uncover. 

\subsection{Understanding Skills of Software Engineers}

To understand what leads to the effectiveness of software engineers that could be missing in AI agents, we draw insights from studies of professional software engineers. There have been multiple studies in literature focused on the attributes of great software engineers. Li et.al \cite{li2020distinguishes} conducted mixed-methods research with experienced software engineers at Microsoft to understand attributes that make a great software engineer. They identified 53 attributes spanning themes like being curious, long-term thinking, making well-considered and informed decisions, sharing knowledge with others, and continuously learning and improving. 

In more specific domains and contexts, Hewner \cite{hewner2010game} explored desired qualifications for new game developers, identifying interpersonal skills—such as the ability to collaborate effectively and set aside personal ego—as crucial alongside core technical competencies. Similarly, Begel et al. \cite{begel2008pair} observed that in pair programming settings, participants favored partners who offered complementary perspectives, demonstrated open-mindedness, and possessed strong communication skills.

We leverage the insights from these studies to inform the design of our expert interviews (\hyperref[sec:study2]{Study 2}). Rather than assuming a direct mapping of human skills to AI agents, we explicitly asked participants to compare their expectations for an AI agent to those for a junior human developer.

Building on these bodies of work, we propose a taxonomy of desirable behaviors for interactive SWE agents. This taxonomy distinguishes itself from previous efforts by addressing multiple dimensions of both the agent's trajectory and its final output, and importantly, it is grounded in user preferences and expectations, distilled from user-defined agent rules in an enterprise setting, and then extended into a framework through expert interviews and prompt analysis in a different work context. 

\section{Study 1: Expected Agent Behavior in Enterprise Software Engineering}
\label{sec:study1}
Our initial effort focused on creating a taxonomy of desirable agent behaviors immediately applicable to an enterprise software engineering context. To identify these behaviors, we performed a qualitative content analysis of user-defined rules\footnote{A general introduction to agent rules can be found at: https://cursor.com/docs/context/rules.} for steering the behavior of a coding agent used at a global technology company. These rules, stored in the codebase as a markdown file, customize the agent's default behavior as a preamble of user prompts.

\subsection{Methods}
In July 2025, we compiled a corpus of 91 agent rule files, each corresponding to a distinct project, from the company's codebase. We started our analysis by manually examining 15 files from our corpus. Two co-authors conducted collaborative and iterative open coding on the initial 15 files from our corpus, which led to a codebook of 15 behavior codes grouped by 4 themes.

To code the rest of the corpus and in anticipation of repeating this analysis as the number of agent rules grow, we tuned an LLM-based rule annotator based on our codebook. The annotator was built on one of the frontier reasoning LLMs. We instructed the model to first segment each file into a set of coherent rules and then annotate each rule with no more than three codes from the codebook.
To mitigate hallucination, the model was given an option to annotate a rule with “TBD” when no code appears to be relevant.  

We evaluated the reliability of the annotator by applying it to 5 randomly selected files from our corpus. Across these 5 files in our validation set, the LLM-based annotator identified 95 unique rules, about 7.2\% of all the rules in our corpus, and achieved a zero hallucination rate, 94.4\% precision, and 91.2\% recall. In this assessment, precision was defined as the proportion of correct rule-code pairs out of all the pairs suggested by the LLM, and recall was defined as the proportion of the former out of all correct rule-code pairs. We consider this performance to be acceptable for our qualitative analysis.

\subsection{\textbf{Findings}}
Our analysis of the agent rules defined by software engineers revealed 4 core expectations and 15 specific behaviors for AI agents, summarized in Table \ref{tab:taxonomy}. In the rest of this section, we describe these expectations in detail, the underlying user needs they reflect, and the interplay of different agent behaviors expected by software engineers. For the purposes of anonymizing the company, we replaced identifying information with a descriptive text enclosed in angle brackets.

\begin{table*}
\centering
\caption{The behavior taxonomy for software engineering agents in an enterprise setting, identified through an analysis of user-defined agent rules.}
\label{tab:taxonomy}
\begin{tabular}{p{0.22\linewidth} p{0.28\linewidth} p{0.4\linewidth}}

\toprule
\textbf{Expectation} & \textbf{Behavior} & \textbf{Definition} \\
\midrule

\multirow{2}{=}{\textbf{1. Adhere to Standards and Processes}} 
 & Follow Established Best Practices & Adhere to established coding patterns for specific languages or frameworks, often by referencing documentation. \\
\cmidrule(l){2-3}
 & Follow Project Workflows and Conventions & Follow project-specific procedures (e.g., dependency management, testing) and conventions (e.g., naming, code organization). \\
\midrule % Line between major groups

\multirow{3}{=}{\textbf{2. Ensure Code Quality and Reliability}} 
 & Maintain Code Style & Follow rules for code formatting, naming conventions, and syntax to ensure consistency. \\
\cmidrule(l){2-3}
 & Write Readable and Maintainable Code & Write code that is easy for humans to understand, modify, and debug by managing comments, complexity, and structure. \\
\cmidrule(l){2-3}
 & Build Robust and Performant Software & Write code that is resilient, well-tested, performant, and observable. \\
\midrule % Line between major groups

\multirow{6}{=}{\textbf{3. Solve Problems Effectively}} 
 & Understand Project Context before Acting & Learn about the project's architecture, components, and core technologies before taking action. \\
\cmidrule(l){2-3}
 & Validate Work Proactively & Verify work by running builds, tests, or other checks to ensure correctness before completion. \\
\cmidrule(l){2-3}
 & Work Incrementally and Iteratively & Break down large tasks into smaller, verifiable steps to reduce complexity and simplify review. \\
\cmidrule(l){2-3}
 & Maintain Task Focus & Limit actions strictly to the user's request, avoiding unrelated changes or refactoring. \\
\cmidrule(l){2-3}
 & Infer Intent from Context & Use the current context (code, request) to resolve ambiguity or fill in missing information proactively. \\
\cmidrule(l){2-3}
 & Learn by Examples & Infer correct patterns and approaches by examining existing code in the project. \\
\midrule % Line between major groups

\multirow{4}{=}{\textbf{4. Collaborate with the User}} 
 & Communicate Effectively & Follow guidelines on how to format output and communicate actions and progress to the user. \\
\cmidrule(l){2-3}
 & Seek Help and Clarification & Ask the user for input or guidance when stuck, unsure, or lacking information. \\
\cmidrule(l){2-3}
 & Plan Collaboratively and Analyze Trade-offs & Create a plan, evaluate different options, and present it to the user for approval before execution. \\
\cmidrule(l){2-3}
 & Learn from Feedback and Past Experiences & Record and apply knowledge from past tasks and mistakes to improve future performance and avoid repeating errors. \\

\bottomrule
\end{tabular}
\end{table*}

\subsubsection{Adhering to Standards and Processes}
A predominant theme was the expectation that the agent must strictly adhere to established standards and project-specific processes. This expectation highlights the "brownfield" nature of software development in a large enterprise where AI-generated code is expected to integrate seamlessly into a mature, existing codebase. Users consider the following two related agent behaviors critical to satisfy this expectation:

\textbf{1. Following Established Best Practices:} Users frequently instructed the agent to "follow," "adhere to," "consult," and "use" pre-existing, widely-accepted guidelines for programming languages and frameworks within the company. These included documents like "CSS Best Practices," "efficient java guide," "Go Style," and the "TypeScript style guide." These rules represent a desire for the agent to align with institutional standards, ensuring its output meets a baseline level of quality and professionalism applicable across multiple projects.

The specificity of these rules can vary across projects. While some rules simply link to online documentation, others underscore key steps directly in the rules file itself, indicating differing degrees of confidence in the agent's capacity to identify and implement guidelines.

\begin{verbatim}
    * Organize your Angular project with one directory per feature.
    * Place common components in a `shared/` directory.
    * Use a single `NgModule` per <build system> package, 
    except for <server platform> Angular applications that must use standalone components.
    * Include a `README.md` file to explain the development environment.
    * Follow the [Angular Style Guide](<URL to the guide>) for more details.
\end{verbatim}

\textbf{2. Following Project Workflows and Conventions:} Beyond general best practices, users specified a multitude of project-specific procedures and conventions. These rules governed local practices such as dependency management, interaction with external systems, test generation and execution, API usage patterns, naming conventions, and code organization.

\begin{verbatim}
    The <package definition> file **MUST** be updated in this specific order:
    1. **Add** After creating your new `\_test.go` file,
    add a `go\_test` target to the `test/<package definition>` file. 
    2. **Verify the test builds:** You **MUST** ensure the test build correctly. 
    If not fix any errors and ensure it builds before proceeding
\end{verbatim}

By providing these rules, users effectively mandated an “orientation” for the AI agent. Their goal is likely to ensure its contributions are not only functional but also conformant, preserving the integrity of the codebase and saving significant downstream costs associated with code review, refactoring, and in the worst case, rework.

\subsubsection{\textbf{Ensure Code Quality and Reliability}}

The agent rules demonstrate a deep concern for multiple facets of code quality, extending far beyond functional correctness. Users instructed the agent to produce code that was not only consistent in style but also readable, maintainable, reliable, and performant. Users articulated these expectations through a variety of directives, which we group into three key areas below.

\textbf{Maintain Code Style:} A significant portion of the rules centered on enforcing stylistic consistency, a critical factor for managing a large-scale, multi-author codebase. These instructions fell into three categories:
\begin{itemize}
    \item \textbf{Rule Application:} Users frequently directed the agent to follow official style guides. These were often supplemented with hyper-specific inline rules for naming conventions, import ordering, or syntax. 
    \item \textbf{Tool Execution:} Adherence was often automated by instructing the agent to execute standard formatting and linting tools as part of its workflow. 
    \item \textbf{Contextual Adaptation:} Some users expected the agent to infer and adapt to the local style of the surrounding code and prioritize local conventions over global ones when they are not aligned. 
\end{itemize}
\begin{verbatim}
    **Follow Local Style:** Adhere to the existing patterns and naming
    conventions of the code you are editing. This takes precedence over general
    style guides.
\end{verbatim}
\textbf{2. Write Readable and Maintainable Code:} Users emphasized the need for AI-generated code that was easy for other humans to understand, modify, and debug. This expectation frames code not as a set of instructions for a machine, but as a form of communication within a development team. One rule demonstrated this principle perfectly:
\begin{verbatim}
    **Write for Others:** Code should be written with the assumption that 
    someone else will read, understand, and maintain it. 
    Your future self is also "someone else".
\end{verbatim}
Specifically, users provided several types of guidance towards code maintainability:

\begin{itemize}
    \item \textbf{High-Level Principles:} Adherence to established software engineering philosophies like DRY ("Don't Repeat Yourself"), "Loose Coupling," and the "Boy Scout Rule" (leave the code cleaner than you found it).
    \item \textbf{Structural Organization:} Instructions to organize code into focused modules, manage dependencies correctly, and, critically, reuse existing utilities rather than creating redundant logic.

\end{itemize}
\begin{verbatim}
    **Helper Functions:** Before implementing common tasks, check for existing
    utilities in <a list of .ts files> and module-specific helpers.
\end{verbatim}
\begin{itemize}
    \item \textbf{Code-Level Simplification:} Directives to write simple, clear code by avoiding deep nesting, simplifying complex conditions, and keeping functions short and focused.
    \item \textbf{Naming and Documentation:} A strong focus on clear, descriptive names for variables and methods, as well as the creation and maintenance of comments and docstrings to provide context for future developers.
\end{itemize}

\textbf{Build Robust and Performant Software:} Finally, users instructed the agent to produce code that was resilient, efficient, and well-tested. They expected the agent to proactively avoid common pitfalls. For example, the rule below guided the agent on memory management:
\begin{verbatim}
    Unsubscribe from observables using `take(1)` 
    or other lifecycle-aware operators to prevent memory leaks.
\end{verbatim}

These rules about code quality communicate a clear user preference that the efficiency gain from agentic coding should not come at the expense of the long-term health and coherence of the codebase. In an enterprise setting, code quality is a multi-faceted concept where users treat functional correctness as a baseline that often goes unmentioned. Instead, users focused their instructions on enforcing the qualities that enable large teams to collaborate effectively over time.

\subsubsection{\textbf{Solve Problems Effectively}}

Beyond adhering to work standards and ensuring code quality, users actively sought to instill effective problem-solving heuristics in the agent. These rules represent a form of procedural knowledge, teaching the agent not just \textit{what} to do, but \textit{how} to approach its tasks like an experienced engineer, such as:

\textbf{Understand Project Context before Acting:} The agent should gather and internalize project-specific information before making changes. This involves reading documentation and examining existing code to understand the project’s architecture, tech stack, and design patterns. 

\textbf{Work Incrementally and Iteratively:} When approaching a large task, the agent should methodically break it down into small, manageable steps. Directives such as "one BUILD file at a time," "small, single-purpose CLs," and making the "smallest possible code change" were common among user-defined rules. 
\begin{verbatim}
    **Incremental Changes:** Make the smallest possible code change, 
    then run tests. Fix any failures before making further changes. 
\end{verbatim}

\textbf{Validate Work Proactively:} Users mandated that the agent take responsibility for the quality of its own output before presenting it to them. The agent was commanded to perform a range of validation actions, including building the project, checking code style, and running tests. 
\begin{verbatim}
    After finishing any changes, always run our unit tests 
    by running `<command>` and iterating until these tests pass. 
\end{verbatim}

\textbf{Maintain Task Focus:} To prevent unintended side effects or scope creep, users imposed strict constraints on the agent's actions. Using imperative phrases like "Only accept," "DO NOT engage," and "Leave... alone," users created clear boundaries, demanding that the agent remain laser-focused on the specific task at hand.

\textbf{Infer Intent from Context:} In a fascinating contrast to the need for strict focus, users also granted the agent a degree of autonomy to resolve ambiguity, especially in low-stakes situations. Rather than halting and asking for clarification, the agent was sometimes expected to use its reasoning capabilities to infer user intent from the available context.

\begin{verbatim}
    If the user does not write any test description, 
    the AI agent should still try inferencing the test case 
    from the test case name, or any other information available. 
\end{verbatim}

\textbf{Learn by Example: }Related to inferring the user’s intent from context, the agent was also expected to reason about the best course of action from examples in the codebase,as shown in the example below:

\begin{verbatim}
    Inspect other tools in the tools/ dir and copy their approaches.
\end{verbatim}

These problem-solving heuristics appear to be driven by a distinct set of underlying user needs:

\begin{itemize}
    \item \textbf{Mitigating Risk:} The emphasis on incremental work—small, atomic commits that can be easily understood, verified, and, crucially, reverted—is a primary risk mitigation strategy.
    \item \textbf{Reducing Human Cognitive Load:} By instructing the agent to validate its own work and fix its own errors, users aim to avoid reviewing half-baked or broken code. This shifts the responsibility for initial quality assurance onto the agent, saving the user's time and attention from performing detailed verification and troubleshooting.
    \item \textbf{Balancing Control and Autonomy:} The rules reveal a fundamental tension in the desired human-agent relationship. On one hand, users impose rigid controls to prevent errors. On the other, they seek to leverage the agent's reasoning capabilities by allowing it to infer intent in certain situations. Such practices reflect an ongoing calibration of the user's mental model of the agent as a tool vs. a partner, reflected in the specific conditions under which the agent was encouraged to exercise more or less autonomy.
\end{itemize}

\subsubsection{\textbf{Collaborate with the User}}
Our analysis reveals that users expect the agent to be more than a passive tool; they envision it as an active collaborator. Users defined protocols for a productive human-agent partnership, focusing on how the agent should communicate, manage uncertainty, formulate plans, and learn over time. Four key collaborative behaviors emerged from our analysis: 

\textbf{Communicate Effectively:} Users required communication that was precise, transparent, and efficient, prioritizing substantive information over conversational pleasantries. This technical communication style aimed to ensure the precision of information, maintain the user’s situational awareness, and facilitate exception management.
\begin{verbatim}
    You are STRICTLY FORBIDDEN from starting your messages with "Great",
    "Certainly", "Okay", "Sure". You should NOT be conversational in your responses, 
    but rather direct and to the point… 
    It is important you be clear and technical in your messages.  
\end{verbatim}

\textbf{Seek Help and Clarification:} The agent was expected to recognize the limits of its own knowledge and capabilities. Under conditions of ambiguity, high-stakes actions, or environmental blockers, it was instructed to pause and defer to the user. Critically, Users also wanted these interruptions to be low-friction, often by having the agent provide suggested answers to make it easy for the user to provide guidance.
\begin{verbatim}
    When you ask a question, provide the user with 2-4 suggested answers 
    based on your question... However if you can use the available tools 
    to avoid having to ask the user questions, you should do so. 
\end{verbatim}
\textbf{Plan Collaboratively and Analyze Trade-offs:} Before executing complex tasks, users required the agent to engage in collaborative planning. This involved creating a plan, evaluating the pros and cons of different approaches, and presenting recommendations for user approval. In some cases, the agent was even expected to exercise critical thinking, challenging suboptimal initial requests and proposing better alternatives with clear justification.

\begin{verbatim}
    Critically evaluate all requests. If a prompt is ambiguous, suboptimal, 
    or follows an anti-pattern, please challenge it and propose a better alternative,
    explaining the trade-offs, especially in the context of data engineering, 
    <a SQL dialet>, and <a query tool>. 
\end{verbatim}

\textbf{Learn from Feedback and Past Experiences:} Users sought to overcome the stateless nature of LLMs by instructing the agent to learn from its mistakes and experiences. While many rules implicitly aimed for this through negative constraints ("never ever do …"), some users attempted to arrange an explicit mechanism for knowledge retention, such as having the agent maintain a "lessons learned" document.

\begin{verbatim}
    At the end of every task, or upon making an error, 
    you **MUST** update the `lessons\_learned.md` file. 
    This is not optional… 
    This process is essential for self-correction and knowledge retention.
\end{verbatim}

The above collaborative protocols defined in agent rules suggest a sophisticated interplay of user needs:

\begin{itemize}
    \item \textbf{Mitigate the Risks:} The demands for planning, transparency, and seeking help are primarily risk-mitigation strategies. Users are concerned about agents performing destructive or incorrect actions without oversight and use these human-in-the-loop checkpoints to maintain control.
    \item \textbf{Reduce Cognitive Load:} Efficient communication and the ability to learn from feedback are aimed at making the agent easier to manage. Conversational fluff is perceived as noise that slows down human guidance, while having to repeat corrections is a source of frustration.
    \item\textbf{Elicit Agency:} Beyond simply preventing errors, some rules aim to leverage the agent's reasoning capabilities. By encouraging the agent to challenge requests, analyze trade-offs, and identify ambiguity, users are prompting the agent to exercise its “agency,” though often within specified limits. 
\end{itemize}

\subsubsection{\textbf{Interplay between Expectations of Agent Behavior}}

Our analysis reveals that the four core behavioral expectations—Adhering to Standards, Ensuring Quality, Solving Problems Effectively, and Collaborating with the User—do not operate in isolation but rather form a tightly integrated system. This interplay highlights a sophisticated user desire for \textit{bounded agency} in AI agents, where autonomy is exercised within defined constraints to achieve a holistic standard of code quality.

Firstly, the expectation to\textit{ adhere to standards and processes} operationalizes quality in an enterprise setting through codified style guides and project conventions. This adherence is not always fully automated; some rules define a collaborative workflow where the agent performs tasks up to a certain point, then hands off to the user for human intervention or specialized tool access, as seen in the two-step process for defining experimentation flags. 
\begin{verbatim}
    So defining the flag id is a two step process:  
    1. Define the flag with a placeholder id which the <tool name> tool 
    can recognize. For example: <code>
    2. Ask the user to run the "<command for an experimentation tool>" command. 
    That tool will replace the placeholder with a unique flag identifier.
\end{verbatim}

Secondly, the heuristics for \textit{solving problems effectively} provide the procedural guardrails necessary to achieve quality safely and efficiently. The emphasis on incremental work and proactive validation creates a tight feedback loop, ensuring quality is integrated from the outset rather than being an afterthought. This desire for a proactive, reasoning agent might seem to contrast with strict process adherence. However, it reflects a nuanced demand for agents to exercise their reasoning capabilities within the specific constraints of the project's context, standards, and problem-solving heuristics. Rules requiring agents to state applicable guidelines and initiate proactive context retrieval before revising a plan illustrate this balance.
\begin{verbatim}
    **after every user prompt your very first output must be**:
    
    1. State which guidelines are applicable with reasoning, 
    and what you will do to follow them.
    2. Initiate `Proactive Context Retrieval`.
    Then based on the context, EXPLICITLY STATE your revised plan 
    and the reasoning behind revision.
\end{verbatim}

Finally, the expectation to \textit{collaborate with the user}, including collaborative planning, trade-off analysis, and learning from feedback, ensures the agent to align the quality of its work with the user's deeper, often under-stated, goals. The heuristics for effective problem-solving, such as "Work Incrementally" and "Validate Work Proactively," are deeply complementary with collaborative expectations. They establish specific protocols that structure human-agent interaction, creating natural breakpoints for user review and guidance.

\section{Study 2: Aspirational Agent Behavior at the Junior Developer Level}
\label{sec:study2}
While the taxonomy of agent behavior developed through \hyperref[sec:study1]{Study 1} is immediately applicable to an enterprise software engineering context, these findings reflect a pragmatic view, shaped by developers’ perceived capabilities and limitations of current LLMs. To ensure the next generations of LLMs progress against the true user need for a highly-intelligent partner in software engineering, we must define a more aspirational target. This requires shifting our focus from current expectations of coding agents to the enduring principles of software engineering, as articulated by experienced practitioners. 

\subsection{\textbf{Methods}}
\begin{figure}[ht]
    \centering
    \includegraphics[width=1\linewidth]{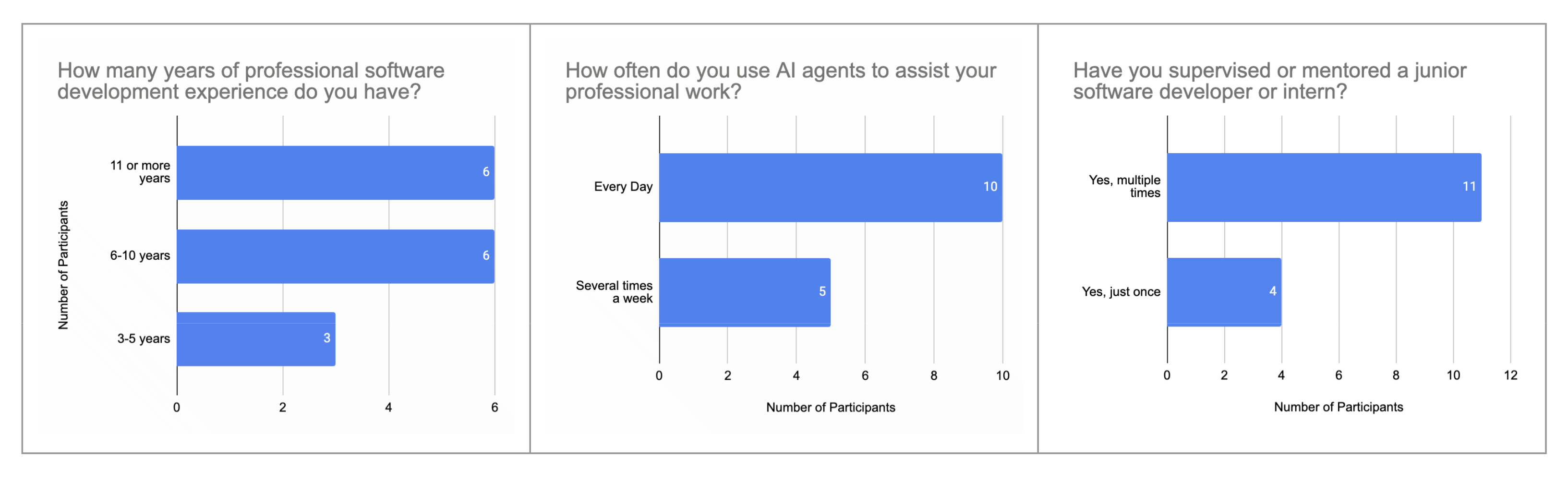}
    \caption{The majority of the 15 developers in our interview study had more than 6 years of professional experience, and the majority use AI agents for their professional work every day. In addition, the majority of our participants supervised or mentored junior software developers multiple times in their respective careers.}
    \label{fig:study2}
    \Description{Three bar charts showing the background information about study participants}
\end{figure}

For the aforementioned purpose, we conducted semi-structured interviews with 15 experienced software engineers who frequently used AI agents for work from the same large technology company (see demographics information in Figure \ref{fig:study2}). We situated our interview protocol in areas of professional growth expected of junior developers as they become seasoned practitioners. Our choice of anchoring our inquiry to expectations of junior developers is an intentional response to the popular framing of AI agents in industry headlines that they could already act like junior software developers\footnote{Goldman Sachs deploys AI software engineer Devin to transform tech roles. https://www.hrkatha.com/news/goldman-sachs-deploys-ai-software-engineer-devin-to-transform-tech-roles/}.

Each interview included two main segments: 1) a mindmapping exercise aimed at eliciting the participant’s view of what constitutes core values and principles in software engineering, often drawing from their own professional journey, and 2) a discussion about how junior developers and AI agents would respectively measure up to these stated ideals. Two of the co-authors conducted thematic analysis on the interview transcripts, informed by the codebook developed from analyzing the agent rules in \hyperref[sec:study1]{Study 1}.

\subsection{\textbf{Findings}}
We found that the expected areas of growth for junior developers are generally aligned with the expectations of (and hopes for) AI agents expressed in agent rules today. For example, participants emphasized the importance of prioritizing code quality over the speed of coding, and believed the key to achieve high code quality is to employ sound problem-solving approaches, such as thinking holistically and making incremental changes, as well as collaborating with others effectively, especially through design and code reviews.

Nonetheless, a crucial difference is that junior developers are expected to take initiatives in their own growth through seeking help and asking questions, observing team workflows and norms, and learning from feedback through both formal reviews and informal discussions with colleagues. For instance, P1 shared a particular mindset he benefited when he was in earlier stages of his career:

\textit{“I think one of the things I've learned is…always make sure that you actually understand the code that you're reading and the technical changes. And don't be afraid to ask other engineers or challenge why the code was written that way. Don't make assumptions that it's already correct or that it should be intuitive.” - P1}

And P10 emphasized the tacit knowledge that needs to be acquired through interacting with team members:

\textit{“There is some sort of team norm (about coding patterns) that needs to be learned over time or through talking to the senior people on the team.” - P10}

In contrast, participants expected AI agents to perform as is, and any seeming improvements of capabilities over the baseline would require active guidance from the user through prompt and context engineering. Therefore, this metacognitive and self-improving capability sets apart human capabilities from AI's, and should be an important consideration when defining and benchmarking against AGI. In the rest of this section, we elaborate on each of the core values junior developers are expected to adopt as they grow, and how the expectations differ when directed towards junior developers and AI agents.

\subsubsection{\textbf{Ensuring Code Quality}}
This is the value that many of our participants considered to be paramount. Participants described “quality” code as code that is correct, readable, maintainable, and thoroughly tested. Participants think of code quality as the social practice of communicating and collaborating with "future engineers" (P1) like "someone who doesn't work at [Company Name] yet" (P9). In that sense, high-quality code—through its readability, comments, and documentation—is a form of asynchronous communication. Tests were described as a "second tier of documentation" (P4) for future engineers.

\textit{“I would say if we don't pay attention to the quality over time, it's gonna bite you back. ... because people come and go, you know, ... some people may have this knowledge and they're gone and we just totally lose that piece. And when we run into something like that, it would take much longer for a team member to figure that out." - P7}

Participants did not anticipate junior developers would appreciate the importance of code quality immediately. They acknowledged that this is an area of initial struggle and eventual growth. Participants expected junior engineers to internalize the mindset of quality first through mentorship, typically iterating on their code via the code review process, and overall building empathy for the reader of their code.

\textit{“There is a tendency for earlier (in their career) developers to get the thing done, and that's it ... their first draft is usually their final draft. ... I think the first thing I try to, you know, work with them is ‘let's iterate.’ Do your own first code review, right? Take off your code author hat and put on a code reviewer hat.” - P6}

On the other hand, an AI agent is still viewed as a tool, despite their potential to become a partner, to which tedious quality tasks (like generating unit tests) can be offloaded immediately and at scale, with no learning curve required.

\textit{“I think AI does a very good job of writing unit tests... normally if I had to write a unit test on my own completely, it'll take like a lot of time” - P10}

\subsubsection{\textbf{Adhering to Standards and Processes}}
Closely related to code quality, participants viewed “internalizing” and adhering to team and organization-wide standards, both explicitly documented best practices as well as implicit “team habits” as an important trait. Adherence to standards enables order and predictability in a large, multi-person codebase, which in turn ensures that the codebase remains maintainable over time.

\textit{“It's not just about the functionality, but also overall readability and whether it aligns with the rest of how [Company Name] writes this piece of code” - P6}

Much like in the case of code quality, participants viewed this as a trait that junior developers would acquire through learning from documentation and more importantly, acting on feedback from code reviews. Junior developers are expected to learn both the documented standards and unwritten, implicit ones embedded in the team’s culture.

“\textit{A lot of it is done in the (code) review process… My tech lead pointed me to the internal [Company Name] C++ documentation to follow like [Company Name] style practice}." - P1

In contrast, they expected AI agents to automate adherence to standards. For example, a coding agent who can rewrite any code to match the standards of surrounding code: 

"\textit{So what I did was I basically used <Agent Name> to say, rewrite what I've done, but in the style of the rest of this file.}" - P9

Still, some participants were disappointed that AI agents were not yet able to learn what standards and rules should be enforced based on a project’s context and an individual’s past work. For example, P2 wanted an agent that can learn from his past code reviews and enforce similar standards in future review requests.

“\textit{The AI could learn from this, [including] 1500 CLs (change lists) I reviewed and what comments I provided, and act on the new CLs.}” - P2.

\subsubsection{\textbf{Solving Problems Effectively}}

Beyond writing good code, a key trait that participants identified as key to being an effective developer was to adopt the right methods for problem solving, including understanding both technical and business requirements, weighing tradeoffs, and considering long-term implications of a solution.

For human developers, participants noted that strategic insight emerges from accumulated experience and a holistic grasp of the system. Junior developers actively cultivate this comprehensive understanding through exploration, diligently "reading team documents," conducting "reasonable research" (P3, P4), and learning incrementally from working with the team over time.

\textit{“Your level of understanding of the code base, right? kind of determines the code you can write... if you know more, like you have more context over larger code base, then you know how to design some part of it” - P5}

For AI agents, any strategic output is attributed to a masterfully crafted prompt. The intelligence and strategic insight originate with the human user; the AI functions merely as a tool to explore that strategy.

\textit{“To me, AI is more of a tool, and so unless the user is thinking about these things first. I wouldn't trust AI to do all of this on behalf of the user yet.” - P14}

Both AI and human developers share a common anti-pattern: an inclination to prematurely execute tasks. Junior developers often face criticism for adhering too rigidly to the problem described on a ticket, failing to grasp the "bigger picture" (P3). Similarly, AI agents exhibit a comparable shortcoming, executing prompts "without thinking about other consequences" (P5). The critical distinction lies in how this narrowness is perceived: for junior developers, it represents a remediable developmental phase, whereas for AI, it is viewed as an inherent technical limitation.

"\textit{...what would make me comfortable is if it (AI) could, um, I think making the code changes is the easier part. If it could also consider and show all the trade-offs it considered. - P6}"

\subsubsection{\textbf{Collaborating with Others}}

Effective collaboration and communication was perceived as an important trait to reduce misalignment and thus wasted work. Soliciting others’ feedback early in one’s problem solving process is considered essential to teamwork: 

“\textit{Code reviews are sort of, that can be too late. ...somebody who has already spent a bunch of effort on a design. And at that point, it's hard to provide input necessarily into some of the choices that are made ... you're going backwards.}” - P12

For junior developers, effective communication was viewed as a developmental journey, requiring them to overcome a reluctance to seek assistance and internalize the understanding that software development is a collaborative endeavor. The expectation was that they should proactively communicate both progress and impediments, avoiding work "in a hole for three days” (P6), echoing the expectation that agents should also operate incrementally and iteratively. In addition, junior developers were expected to demonstrate effective communication by adding adequate context in CL descriptions to aid reviewers, as well as create more durable artifacts like documentation to enable asynchronous communication. 

For AI Agents, proactive communication was required to transform its “black box” (P14) nature and build trust. Participants expected that AI agents need to be explicitly designed for interactive dialog and transparent communication. They expect agents to explain the reasoning behind suggestions and changes as well as proactively identify knowledge gaps and ask clarifying questions, even presenting tradeoffs. 

\textit{“}\textit{I would expect more of a conversation like [interaction] - }\textit{here's the problem, here's [how] I would test it... but there was no such conversation [with AI].” - P1}

\textit{“It needs to be able to figure out when it doesn't have enough information and ask... It's gonna be more of an interaction versus like correction... I expected it to say something. ‘Oh, like I have these two possible scenarios... these are the trade offs for left and right... what would you recommend?’” - P3}

\subsubsection{\textbf{Learn from Feedback and Past Experiences}}

Closely related to being an effective collaborator, is being able to learn from feedback and past experiences. This proactive effort on part of the junior developer or AI greatly improves continuity and efficiency: the senior developer doesn’t have to explain the context again, provide the same feedback again, or fix the same bugs as before.

Effective collaboration is intrinsically linked to the ability to learn from feedback and prior interactions. Exhibiting this behavior, whether by a junior developer or an AI agent, enhances efficiency and outcomes of teamwork, alleviating the need for senior developers to repeatedly re-explain context, reiterate feedback, or re-address recurring issues.

For junior developers, participants perceived learning from feedback as an active engagement with the code review process, necessitating an openness to constructive criticism ("\textit{hard for your ego to just say, 'Oh, good idea'}" - P3). For Agents, learning from feedback was closely related to retaining context and personalizing suggestions based on a user's past actions.

Participants in the study acknowledged the potential for AI to be more than a just a tool and an effective partner, given it has already been trained on the organization’s code base, but lamented that AI’s stateless manner, as in its inability to remember the "conversation leading up to" the current moment and learn from past interactions, could make it a forgettable partner.

\textit{“...I would make changes here just fixing bugs. Then if I added any other prompt. It would reproduce the old bug as opposed to taking into account the fact that I changed the code." - P13}

\subsection{\textbf{Introducing the Time Horizon Axis of the CAB Framework}}

\begin{figure}[ht]
    \centering
    \includegraphics[width=1\linewidth]{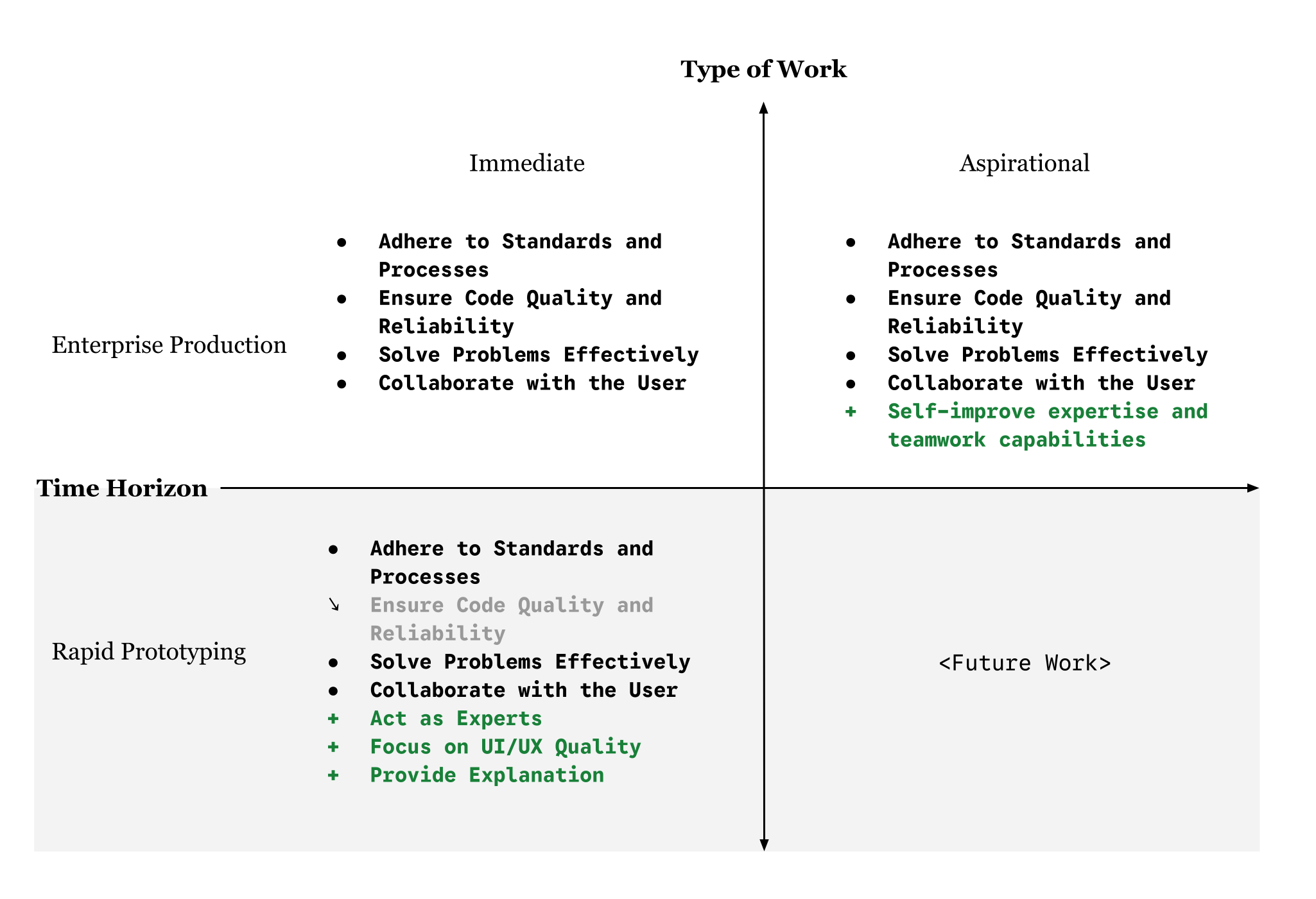}
    \caption{The Context-Adaptive Behavior (CAB) Framework extends the taxonomy of desirable agent behaviors developed for an enterprise setting along two empirically-derived axes. The \textit{Time Horizon} axis maps the evolution from current needs to future ideals. The \textit{Type of Work} axis reveals how expectations change when shifting from production-grade engineering to exploratory development. Text in green with a plus sign indicates expected behaviors unique in its quadrant. Text in gray with a downward arrow represents an expectation less prominent compared to the “Enterprise Production” base line.}
    \label{fig:cab_framework}
    \Description{A 2 by 2 diagram illustrating the Context-Adaptive Behavior Framework}
\end{figure}

To sum up, we found that while engineers expect both junior developers and AI agents to adhere to the same core values of software engineering, the crucial distinction lies in the expectation of growth. Junior developers are expected to be proactive, self-directed learners who internalize feedback and improve over time, whereas AI agents are perceived as static tools that require explicit, continuous guidance from the user to perform better.

Based on these findings, we propose the Time Horizon axis of our Context-Adaptive Behavior Framework (see the top two quadrants in Figure \ref{fig:cab_framework}), extending the near-term user expectations of agent behavior captured in the taxonomy presented in \hyperref[sec:study1]{Study 1}. This axis illustrates the progression from a competent but dependent tool to a junior partner with both a growth mindset and an aptitude to improve continuously. The "Aspirational" end of the axis is thus defined by the emergence of metacognitive capabilities: the ability to learn from experience, proactively seek knowledge, and demonstrate a deep, internalized understanding of a project's context and standards, much like a human developer is expected to.

In the following section, we describe how we derived the Type of Work axis of the framework through examining user prompts for an AI-powered prototyping tool.

\section{Study 3: Expected Agent Behavior for AI-Driven Prototyping}
\label{sec:study3}

To investigate how agent behavior ought to adapt across diverse work contexts, we analyzed user prompts within an AI-driven application prototyping environment. This prototyping tool, developed by the same technology company involved in our initial two studies, attracts developers who primarily explore new application ideas through conversational interactions with an agent. The technical backgrounds of these developers are more varied than the enterprise developers involved in our first two studies, ranging from seasoned full-stack engineers to hobbyists with minimal coding experience, all of whom aim to rapidly iterate on their concepts through conversing with a prototyping agent. This study, therefore, aims to define the "Type of Work" axis within the Context-Adaptive Behavior (CAB) Framework, shedding light on how expected AI agent behaviors both generalize and evolve as tasks shift from production-grade engineering to UI-focused new product exploration.

\subsection{\textbf{Method}}
Users in this prototyping environment instruct the agent through iterative, conversational prompts. While a mechanism for defining agent rules does exist, it is cumbersome and rarely used in this context. Therefore, prompts sent to the prototyping agent served as the primary source for understanding this user group's expectations of agent behavior.

We took a random sample of 1,000 users who had engaged with the agent at least 10 times in July 2025. From this group, we randomly selected 1,000 prompts. We then used a multi-step, LLM-assisted workflow (using one of the frontier reasoning LLMs) to analyze and categorize these prompts.
\begin{enumerate}
    \item \textbf{Filtering}: First, the LLM classified whether each prompt described a preferred agent behavior (e.g., a process or an approach) rather than just a direct task instruction. This yielded a core dataset of 488 prompts.
    \item \textbf{Clustering}: Next, we used the LLM to cluster these 488 prompts by distinct behavioral expectations of the user. Two co-authors manually reviewed the clustering results and refined them into 13 themes.
    \item \textbf{Classification}: Finally, we used the 13 refined themes to classify the 488 prompts. One of the co-authors independently annotated a subset of prompts (n=25) to validate the LLM’s classification accuracy and found a strong F1-score of 83\% (Precision: 81\%; Recall: 85\%).
\end{enumerate}

\subsection{\textbf{Findings}}
In the rest of this section, we focus on the similarities and differences between the expected agent behaviors implied in user prompts and those found in the agent rules analysis.

\subsubsection{\textbf{Expectations Similar between Enterprise Software Development and Rapid Prototyping}}

There’s a set of behavioral expectations that seem to apply to both settings, though often expressed differently:

\begin{itemize}
    \item \textbf{Adhere to Standards and Processes: }Users often instruct the agent to follow best practices at a high level, such as, “\textit{Make sure to follow best practices for data and web application security}” or “\textit{Ensure consistent naming conventions, data types, and required fields for all data points that will be used for analytics.}” Users also specify how to perform a specific task and provide step-by-step plans. For example, users would enter “\textit{now perform the steps that you came up with in your last answer}” to remind the tool of previously established plans, or enter instructions themselves, such as "\textit{here are the exact steps you should take to troubleshoot and resolve this issue.}"
    
    \item \textbf{Ensure Code Quality: }In the prototyping setting, the range of rules imposed for prototyping outcome ranged from strict (e.g., “\textit{Layout must follow best practices for mobile-first design: Top padding, clean safe margins, Max readability, Responsive layout}”) to open-ended (e.g., “\textit{build a user friendly app}”, “\textit{review the entire codebase again from the perspective of a professional coder and search for any inefficiencies, redundancies or unprofessional code that could be improved}”).
    \item \textbf{Collaborate with the User:} In both types of work, users expect the agent to engage in collaborative planning. This is evident in prompts that invite the prototyping agent to brainstorm, assess task difficulty, or ask for clarification, such as:
    \begin{itemize}
        \item “\textit{what else do you think we should add?}”
        \item “\textit{Good idea, don't make it too dominant but I like the thinking here. You can implement this strategy in multiple phases, if it means for a better output?}”
        \item “\textit{If something is unclear, request clarification rather than assuming.}”
    \end{itemize}
    \item \textbf{Solve Problems based on Context:} Like developers in the enterprise setting, users of this prototyping agent expected AI to acquire and apply knowledge from its context. For example, some users asked the agent to look back into the conversation history, “\textit{Examine our recent prompt history and think very deeply about a solution that isn't something we already tried.}” Others requested the agent to review the project context, “\textit{based on the context of our project as defined in the docs/rules.md, docs/schemas.md, readme.md and openapi.yaml files, how would you make this plan more robust?}”
    \item \textbf{Validate Work Proactively and Learn from Feedback: }When users experience regression and persistent failures, they try to address it with a reactive prompt with expression of frustration (e.g., \textit{"why do you always make the same mistake?}"). The underlying expectation is to “validate work proactively,” and  “learn from feedback and past experience.”
\end{itemize}

\subsubsection{\textbf{Expectations More Prominent for Rapid Prototyping}}

We also observed a set of user expectations that appear to be more prominent in the rapid prototyping setting than enterprise software engineering: 

\begin{itemize}
    \item \textbf{Act as Experts: }A common pattern in prompting the prototyping agent is the direction to assume the role of an expert. Examples include: “\textit{You are the best UI designer and your designs are loved by GenZ}”, “\textit{You are a skilled game designer with extensive experience in creating immersive 3D car racing games.}”, and "\textit{You are an expert full-stack developer.”} This prompting style represents the most abstract form of behavioral guidance, suggesting some users may lack the expertise to provide concrete guidance. Invoking an expert persona might have been a workaround to elicit desirable behaviors they are not able to articulate.
    \item \textbf{Focus on UI/UX Quality: }While agent rules in the enterprise center on code health (reliability, maintainability, performance), the prompts for prototyping show a strong focus on the quality of visual and UI design. Users provide high-level guidance such as "\textit{make it significantly more modern and minimalist}" or "\textit{like Apple's design,}" in addition to specific task instructions (\textit{“change the purple color to a dark navy blue”}).
    \item \textbf{Provide Explanation: }Users frequently ask the agent to explain and clarify the agent’s plans and actions. This happens when they are in a collaborative planning mode (e.g., \textit{“Think deeply about the implementation, and explain the approach.}”) and when they want to gain insights into the agent’s decisions (\textit{“so help me understand the logic, if is not logged in it will be redirected to homepage?”}). Some requests for explanation also reflect a desire to learn. For example, users would ask, “\textit{how can that be implemented, or is that a possible upgrade to add later on}” or \textit{“It's working, explain how you went from genkit to fetch? What's the difference?”}. This signals varying levels of technical expertise of users and a pedagogical need not as present in the enterprise setting.
\end{itemize}

\subsubsection{\textbf{Expectations Less Prominent in Rapid Prototyping}}

\textbf{Ensure Code Quality: }User prompts for the prototyping agent included fewer explicit requirements on code quality than the enterprise agent rules did. This is not surprising given the differences between the two settings. Nonetheless, users in this setting still have quality concerns over the agent’s work. While their primary focus is on the UI/UX quality (e.g., “\textit{Layout must follow best practices for mobile-first design: Top padding, clean safe margins, Max readability, Responsive layout}”), some users requested the agent to improve the health of their code: “\textit{review the entire codebase again from the perspective of a professional coder and search for any inefficiencies, redundancies or unprofessional code that could be improved.}”

\subsection{\textbf{Reflecting on Our Observations}}

The analysis of user prompts for the rapid prototyping agent confirms that the core behavioral expectations codified in the enterprise agent rules are broadly applicable. However, these expectations manifest with varying levels of strength and specificity across these two distinct settings of software development. We updated the top-right quadrant of the CAB Framework (Figure \ref{fig:cab_framework}) accordingly to reflect these nuanced differences.

Two primary factors seem to be behind the differences in desirable agent behavior we observed. To begin with, the user base for rapid prototyping is more diverse, encompassing both professional and casual developers. This variance in user expertise might have led to different ways they articulate their expectations of the agent. The underlying purpose of the code also significantly influences the prompt language. While a fundamental desire for quality persists, work with the prototyping agent often prioritizes idea exploration, rapid iteration, and immediate functionality over high-stakes production deployments. Consequently, users employ less formal and more indirect expressions to convey their underlying expectations. This observation highlights the necessity for an agent evaluation framework to recognize and adapt to these implicit, informally articulated user needs.

\section{Discussions}

\subsection{\textbf{The Evolving Mental Model of AI Agents}}
Our work highlights an evolving mental model of AI agents in software engineering: the simultaneous perception of AI as both a powerful tool and a potential collaborative partner. This perception manifests in several key ways, influencing how users interact with and evaluate these agents.

First, the varying abstraction levels of agent rules reveal this evolution of mental models directly. In \hyperref[sec:study1]{Study 1}, users provided both high-level principles and highly specific, step-by-step directives. The former suggests a desire for a partner capable of autonomous reasoning and adherence to broad guidelines, while the latter treats the agent as a tool that requires precise instructions to execute tasks. This duality underscores the current stage of AI development, where agents demonstrate advanced inference capabilities but still require explicit, step-by-step guidance in certain contexts.

Furthermore, our findings from \hyperref[sec:study2]{Study 2} suggest that developers appreciate the fact that AI agents often possess superhuman breadth of knowledge, but at the same time they are acutely aware of their shortcomings in areas such as self-improvement and learning on the job, which is in contrast to their expectations of other developers, even junior ones. 

Finally, the user's perception of their own expertise likely shapes their perceived relationship with the AI. In \hyperref[sec:study3]{Study 3}, when users perceive themselves to be less knowledgeable than the AI agent, they tend to adopt a "partner" mindset, expecting the AI to act as a tutor or expert, expressed in role-playing commands (e.g., \textit{"You are the best UI designer and your designs are loved by GenZ"}). Conversely, when users perceive a significant expertise advantage of their own over the AI, as shown in \hyperref[sec:study1]{Study 1}, they revert to using the AI more as a sophisticated tool or a junior partner at the best. This suggests that the "tool vs. partner" dynamic is not inherent to the AI itself, but rather a fluid construct influenced by the user's context and self-perception.

\subsection{Implications for Improving Agent Evaluation}
To effectively evaluate the progression from AI-as-a-tool to AI-as-a-partner, our taxonomy of desirable agent behaviors and the Context-Adaptive Behavior (CAB) Framework offer a human-centered foundation for enhancing our toolbox of evaluation techniques. 

\textbf{Qualitative Assessment of Agent Behavior}: Our framework enables a more nuanced qualitative assessment of agent behavior. By leveraging techniques like "Report Cards" \cite{yang2024report}, each behavior within our taxonomy can be treated as a "Skill." Human subject-matter experts or well-calibrated LLM-as-a-Judge can then assess agent trajectories and code output against specific guidelines for each skill, providing a grade and qualitative assessment. This moves beyond functional correctness to capture the interactive and collaborative qualities essential for partnership. 

\textbf{In-Product User Feedback}: The behavioral taxonomy can directly structure in-product feedback interfaces, leading to more precise and actionable insights. If a user has defined agent rules, the feedback UI can be tailored to align the general taxonomy with their specific rules, focusing on behaviors relevant to the current session. This allows for targeted feedback on how well the agent is performing as a partner in real-world scenarios.

\textbf{Up-leveling Error Analysis to Behavioral Analysis}: Our taxonomy provides a crucial foundation for future work in up-leveling error analysis to behavioral analysis. While this paper introduces the behavioral attributes, developing a hierarchical analytical framework that can attribute high-level behavioral deficits to low-level system errors and limitations is a significant next step. This will allow for a more comprehensive understanding of agent shortcomings and guide targeted improvements, moving beyond mechanistic failures to human-centric behavioral analysis.

\subsection{Implications for Improving Agent Design}
Our analysis also highlights key areas for improving agent behavior to foster more effective human-AI partnerships.

\textbf{Memory and Personalization}: A significant limitation of current AI agents, as observed in our studies, is their often stateless nature. This hinders their ability to learn from past interactions and adapt to individual user preferences, leading to frustration and the need for repeated corrections. Developing robust mechanisms for persistent memory and personalized learning will be essential for agents to evolve into truly adaptive and trustworthy partners. 

\textbf{Communication and Collaboration Tooling}: Agent-to-user communication is often limited to a chat box interface. To truly act as partners, agents need to be better equipped to communicate their reasoning, explain complex decisions, and provide transparent updates on their progress, but also to seek clarification when uncertain, both in synchronous and asynchronous collaborations with humans.This raises the question of providing agents with more access to existing human-to-human communication and collaborative work infrastructure, such as emails, issue trackers, and online collaborative documents, to enable more seamless integration into team workflows.

\textbf{Meta-Cognition Support}: Empowering agents with meta-cognitive capabilities will elevate them from task executors to strategic partners. Key aspects of meta-cognition support for agents could include:

\begin{itemize}
    \item \textbf{Plan}: Before starting a task, the agent can strategize the best way to approach it according to knowledge about its own strengths and weaknesses, based on prior evaluation results, as well as the complexity of the task.

    \item \textbf{Monitor}: While performing the task, the agent can pause and assess its own execution traces and identify any difficulties or gaps in its knowledge.
    
    \item \textbf{Evaluate}: After completing the task, the agent can reflect on its performance and save the results of such a self-assessment in the agent's memory system, leveraged for future tasks.
\end{itemize}

\textbf{“Over-the-Shoulder” Learning}: To match the self-improving capacity of junior developers, as Study 2 suggests, AI agents need to learn from observing how their human partners perform tasks, in a manner similar to “over-the-shoulder learning”\cite{twidale2013over}. This involves developing mechanisms for agents to passively human workflows, identify patterns, and internalize best practices, thereby acquiring tacit knowledge that is often difficult to codify in explicit rules.

\subsection{Implications for Supporting Human Supervisors of  AI Agents}
Finally, supporting users who supervise AI agents is a critical, often overlooked, aspect of fostering effective human-AI collaboration. This includes developing:

\textbf{Agent Rule Validators:} Tools that help users iteratively test and refine their agent rules against their own use cases can be considered as project-specific agent evaluations. This capability enables users to fine-tune agent behavior for optimal performance within their unique workflows and project requirements, ensuring that agents act as effective and reliable partners.

\textbf{Agent Rule Discovery and Adaptation:} Enabling agent rule discovery can empower users to learn successful strategies from their peers and subsequently adapt validated rules to their unique contexts. This fosters a community of practice for agent customization.

\textbf{Agent Profile:} Providing users with clear "agent profiles" that detail an agent's evaluated behavioral strengths and weaknesses can help users calibrate their trust and focus their rule authoring effort on specific types of agent behavior that require explicit guidance

\section{Limitations}
Our research, while offering valuable insights, has certain limitations. Firstly, the data for both \hyperref[sec:study1]{Study 1} (agent rules) and \hyperref[sec:study2]{Study 2} (interview data) were collected exclusively from a single large technology company. While this allows for in-depth analysis within a specific context, it may limit the generalizability of our findings to other organizational settings, company cultures, or industries. Future research could benefit from a broader sample across diverse companies to validate and expand upon our observations.

Secondly, the "Type of Work" axis in our Context-Adaptive Behavior (CAB) Framework is not exhaustive. We specifically focused on two distinct types of software engineering work: enterprise software development and rapid prototyping. While these two represent significantly different purposes and user expertise levels, thus effectively demonstrating the generalizability of our behavioral taxonomy for software engineering agents and the utility of the CAB framework, other types of software engineering work (e.g., embedded systems development, data science projects, game development) were not included. Exploring these additional work types in future studies would further refine and strengthen the comprehensive applicability of our framework.

\section{Conclusions}
Our work addresses a critical evaluation gap in the evolving landscape of AI agents for software engineering. Our core contributions include a foundational taxonomy of desirable AI agent behaviors for enterprise software engineering, derived from a qualitative analysis of 91 agent configurations, and the emerging Context-Adaptive Behavior (CAB) Framework for extending the taxonomy to different Types of Work and Time Horizons of AI advancements. The taxonomy defines four key expectations of agent behavior: \textit{Adhere to Standards and Processes}, \textit{Ensure Code Quality and Reliability}, \textit{Solving Problems Effectively}, and \textit{Collaborating with the User}, providing a systematic vocabulary for understanding effective agent partnership. Together, our taxonomy and framework provide a robust, human-centered foundation for assessing true collaborative intelligence in AI agents, guiding future research and development towards more effective and trustworthy human-AI partnerships.

\section{GenAI Usage Disclosure}
Throughout this paper, GenAI tools were utilized for copy editing. Additionally, these tools assisted with data analysis, as detailed in the pertinent study sections. The co-authors have reviewed the GenAI output for accuracy and deem any potential errors acceptable given the qualitative nature of our analyses.


\bibliographystyle{ACM-Reference-Format}
\bibliography{references}

@inproceedings{vaithilingam2022expectation,
  title = {Expectation vs. experience: Evaluating the usability of code generation tools powered by large language models},
  author = {Vaithilingam, Priyan and Zhang, Tianyi and Glassman, Elena L.},
  booktitle = {CHI Conference on Human Factors in Computing Systems Extended Abstracts},
  pages = {1--7},
  year = {2022},
  doi = {10.1145/3491101.3519665},
  url = {https://doi.org/10.1145/3491101.3519665}
}

@inproceedings{kang2023large,
  title = {Large language models are few-shot testers: Exploring LLM-based general bug reproduction},
  author = {Kang, Sungmin and Yoon, Juyeon and Yoo, Shin},
  booktitle = {2023 IEEE/ACM 45th International Conference on Software Engineering (ICSE)},
  pages = {2312--2323},
  year = {2023},
  organization = {IEEE},
  doi = {10.1109/ICSE48619.2023.00194},
  url = {https://doi.org/10.1109/ICSE48619.2023.00194}
}

@inproceedings{nikolov2025google,
  title = {How is Google using AI for internal code migrations?},
  author = {Nikolov, Stoyan and Codecasa, Daniele and Sj{\"o}vall, Anna and Tabachnyk, Maxim and Chandra, Satish and Taneja, Siddharth and Ziftci, Celal},
  booktitle = {2025 IEEE/ACM 47th International Conference on Software Engineering: Software Engineering in Practice (ICSE-SEIP)},
  year = {2025},
  note = {conference paper; arXiv version available},
  url = {https://arxiv.org/abs/2501.06972}
}

@article{meng2024empirical,
  title={An empirical study on llm-based agents for automated bug fixing},
  author={Meng, Xiangxin and Ma, Zexiong and Gao, Pengfei and Peng, Chao},
  journal={arXiv preprint arXiv:2411.10213},
  year={2024},
  url = {https://arxiv.org/abs/2411.10213}
}

@inproceedings{rondon2025evaluating,
  title = {Evaluating agent-based program repair at Google},
  author = {Rondon, Pat and Wei, Renyao and Cambronero, Jos{\'e} and Cito, J{\"u}rgen and Sun, Aaron and Sanyam, Siddhant and Tufano, Michele and Chandra, Satish},
  booktitle = {2025 IEEE/ACM 47th International Conference on Software Engineering: Software Engineering in Practice (ICSE-SEIP)},
  year = {2025},
  doi = {10.1109/ICSE-SEIP66354.2025.00038},
  url = {https://arxiv.org/abs/2501.07531},
  note = {arXiv preprint also available; accepted/presented at ICSE-SEIP 2025.}
}

@article{kumar2025sharp,
  title={Sharp Tools: How Developers Wield Agentic AI in Real Software Engineering Tasks},
  author={Kumar, Aayush and Bajpai, Yasharth and Gulwani, Sumit and Soares, Gustavo and Murphy-Hill, Emerson},
  journal={arXiv e-prints},
  pages={arXiv:2506.12347},
  year={2025},
  url = {https://arxiv.org/abs/2506.12347},
  note = {Microsoft Research publication / arXiv — no separate peer-reviewed venue found}
}

@article{liu2024large,
  title={Large language model-based agents for software engineering: A survey},
  author={Liu, Junwei and Wang, Kaixin and Chen, Yixuan and Peng, Xin and Chen, Zhenpeng and Zhang, Lingming and Lou, Yiling},
  journal={arXiv preprint arXiv:2409.02977},
  year={2024},
  url = {https://arxiv.org/abs/2409.02977}
}

@inproceedings{jimenez2024swebench,
  title = {{SWE}-bench: Can Language Models Resolve Real-world GitHub Issues?},
  author = {Jimenez, Carlos E. and Yang, John and Wettig, Alexander and Yao, Shunyu and Pei, Kexin and Press, Ofir and Narasimhan, Karthik},
  booktitle = {The Twelfth International Conference on Learning Representations (ICLR 2024)},
  year = {2024},
  url = {https://openreview.net/forum?id=VTF8yNQM66},
  note = {original arXiv:2310.06770}
}

@inproceedings{jain2025livecodebench,
  title={LiveCodeBench: Holistic and Contamination Free Evaluation of Large Language Models for Code},
  author={Jain, Naman and Han, King and Gu, Alex and Li, Wen-Ding and Yan, Fanjia and Zhang, Tianjun and Wang, Sida and Solar-Lezama, Armando and Sen, Koushik and Stoica, Ion},
  booktitle = {The Thirteenth International Conference on Learning Representations (ICLR 2025)},
  year = {2025},
  url = {https://openreview.net/forum?id=chfJJYC3iL},
  note = {original arXiv:2403.07974}
}

@misc{AiderPolyglotBenchmark,
    author = {Aider blog},
    title = {o1 tops aider’s new polyglot leaderboard: The polyglot benchmark},
    year = {2024},
    month = {December},
    day = {21},
    url = {https://aider.chat/2024/12/21/polyglot.html#the-polyglot-benchmark},
    urldate = {2025-10-03}
}

@ARTICLE{taylor2025software,
    author={Taylor, Claire and Huber, Marie and Ma, Qiao and Plaza, Rayven and Chang, Alison and Chen, Jie},
    journal={ IEEE Software },
    title={ Software Development Is a Team Sport },
    year={2025},
    volume={42},
    number={03},
    ISSN={1937-4194},
    pages={13-17},
    doi={10.1109/MS.2025.3539403},
    url = {https://doi.ieeecomputersociety.org/10.1109/MS.2025.3539403},
    publisher={IEEE Computer Society},
    address={Los Alamitos, CA, USA},
    month=may
}

@article{deshpande2025trail,
    title={TRAIL: Trace Reasoning and Agentic Issue Localization},
    author={Deshpande, Darshan and Gangal, Varun and Mehta, Hersh and Krishnan, Jitin and Kannappan, Anand and Qian, Rebecca},
    journal={arXiv preprint arXiv:2505.08638},
    year={2025},
    url = {https://arxiv.org/abs/2505.08638}
}

@inproceedings{cemri2025multi,
    title={Why Do Multiagent Systems Fail?},
    author={Melissa Z Pan and Mert Cemri and Lakshya A Agrawal and Shuyi Yang and Bhavya Chopra and Rishabh Tiwari and Kurt Keutzer and Aditya Parameswaran and Kannan Ramchandran and Dan Klein and Joseph E. Gonzalez and Matei Zaharia and Ion Stoica},
    booktitle={ICLR 2025 Workshop on Building Trust in Language Models and Applications},
    year={2025},
    url={https://openreview.net/forum?id=wM521FqPvI}
}

@article{chen2021evaluating,
    title={Evaluating large language models trained on code},
    author={Chen, Mark and Tworek, Jerry and Jun, Heewoo and Yuan, Qiming and Pinto, Henrique P. de O. and Kaplan, Jared and Edwards, Harri and Burda, Yuri and Joseph, Nicholas and Brockman, Greg and others},
    journal={arXiv preprint arXiv:2107.03374},
    year={2021},
    url = {https://arxiv.org/abs/2107.03374}
}

@article{austin2021program,
    title={Program synthesis with large language models},
    author={Austin, Jacob and Odena, Augustus and Nye, Maxwell and Bosma, Maarten and Michalewski, Henryk and Dohan, David and Jiang, Ellen and Cai, Carrie and Terry, Michael and Le, Quoc and others},
    journal={arXiv preprint arXiv:2108.07732},
    year={2021},
    url = {https://arxiv.org/abs/2108.07732}
}

@inproceedings{miserendino2025swelancer,
    title={SWE-Lancer: Can Frontier LLMs Earn \$1 Million from Real-World Freelance Software Engineering?},
    author={Miserendino, Samuel and Wang, Michele and Patwardhan, Tejal and Heidecke, Johannes},
    booktitle = {Proceedings of the 2025 International Conference on Machine Learning (ICML 2025)},
    year={2025},
    note = {ICML 2025 (oral); arXiv:2502.12115},
    url = {https://arxiv.org/abs/2502.12115}
}

@article{usrey1996dimensions,
    title={The dimensions of software quality},
    author={Usrey, Michael W and Dooley, Kevin J},
    journal={Quality Management Journal},
    volume={3},
    number={3},
    pages={67--86},
    year={1996},
    publisher={Taylor \& Francis}
}

@inproceedings{takerngsaksiri2025code,
    title={Code readability in the age of large language models: An industrial case study from Atlassian},
    author={Takerngsaksiri, Wannita and Tantithamthavorn, Chakkrit and Fu, Micheal and Pasuksmit, Jirat and Chen, Kun and Wu, Ming},
    booktitle = {IEEE International Conference on Software Maintenance and Evolution (ICSME) 2025, Industry Track},
    year={2025},
    url = {https://arxiv.org/abs/2501.11264},
    note = {arXiv preprint and ICSME 2025 Industry Track}
}

@article{zheng2024beyond,
    title={Beyond correctness: Benchmarking multi-dimensional code generation for large language models},
    author={Zheng, Jiasheng and Cao, Boxi and Ma, Zhengzhao and Pan, Ruotong and Lin, Hongyu and Lu, Yaojie and Han, Xianpei and Sun, Le},
    journal={arXiv preprint arXiv:2407.11470},
    year={2024},
    url = {https://arxiv.org/abs/2407.11470}
}

@article{bouzenia2025understanding,
    title={Understanding Software Engineering Agents: A Study of Thought-Action-Result Trajectories},
    author={Bouzenia, Islem and Pradel, Michael},
    journal={arXiv preprint arXiv:2506.18824},
    year={2025},
    url = {https://arxiv.org/abs/2506.18824}
}

@article{li2025counselbench,
    title={CounselBench: A Large-Scale Expert Evaluation and Adversarial Benchmark of Large Language Models in Mental Health Counseling},
    author={Li, Yahan and Yao, Jifan and Bunyi, John Bosco S and Frank, Adam C and Hwang, Angel and Liu, Ruishan},
    journal={arXiv preprint arXiv:2506.08584},
    year={2025},
    url = {https://arxiv.org/abs/2506.08584}
}

@article{ayers2023comparing,
    title={Comparing physician and artificial intelligence chatbot responses to patient questions posted to a public social media forum},
    author={Ayers, John W and Poliak, Adam and Dredze, Mark and Leas, Eric C and Zhu, Zechariah and Kelley, Jessica B and Faix, Dennis J and Goodman, Aaron M and Longhurst, Christopher A and Hogarth, Michael and others},
    journal={JAMA Internal Medicine},
    volume={183},
    number={6},
    pages={589--596},
    year={2023},
    doi = {10.1001/jamainternmed.2023.1838},
    publisher={American Medical Association}
}

@article{maurya2024unifying,
    title={Unifying AI tutor evaluation: An evaluation taxonomy for pedagogical ability assessment of LLM-powered AI tutors},
    author={Maurya, Kaushal Kumar and Srivatsa, KV and Petukhova, Kseniia and Kochmar, Ekaterina},
    journal={arXiv preprint arXiv:2412.09416},
    year={2024},
    url = {https://arxiv.org/abs/2412.09416}
}

@article{shi2025educationq,
    title={EducationQ: Evaluating LLMs' Teaching Capabilities Through Multi-Agent Dialogue Framework},
    author={Shi, Yao and Liang, Rongkeng and Xu, Yong},
    journal={arXiv preprint arXiv:2504.14928},
    year={2025},
    url = {https://arxiv.org/abs/2504.14928}
}

@misc{jurenka2024towards,
      title={Towards Responsible Development of Generative AI for Education: An Evaluation-Driven Approach}, 
      author={Irina Jurenka and Markus Kunesch and Kevin R. McKee and Daniel Gillick and Shaojian Zhu and Sara Wiltberger and Shubham Milind Phal and Katherine Hermann and Daniel Kasenberg and Avishkar Bhoopchand and Ankit Anand and Miruna Pîslar and Stephanie Chan and Lisa Wang and Jennifer She and Parsa Mahmoudieh and Aliya Rysbek and Wei-Jen Ko and Andrea Huber and Brett Wiltshire and Gal Elidan and Roni Rabin and Jasmin Rubinovitz and Amit Pitaru and Mac McAllister and Julia Wilkowski and David Choi and Roee Engelberg and Lidan Hackmon and Adva Levin and Rachel Griffin and Michael Sears and Filip Bar and Mia Mesar and Mana Jabbour and Arslan Chaudhry and James Cohan and Sridhar Thiagarajan and Nir Levine and Ben Brown and Dilan Gorur and Svetlana Grant and Rachel Hashimshoni and Laura Weidinger and Jieru Hu and Dawn Chen and Kuba Dolecki and Canfer Akbulut and Maxwell Bileschi and Laura Culp and Wen-Xin Dong and Nahema Marchal and Kelsie Van Deman and Hema Bajaj Misra and Michael Duah and Moran Ambar and Avi Caciularu and Sandra Lefdal and Chris Summerfield and James An and Pierre-Alexandre Kamienny and Abhinit Mohdi and Theofilos Strinopoulous and Annie Hale and Wayne Anderson and Luis C. Cobo and Niv Efron and Muktha Ananda and Shakir Mohamed and Maureen Heymans and Zoubin Ghahramani and Yossi Matias and Ben Gomes and Lila Ibrahim},
      year={2024},
      eprint={2407.12687},
      archivePrefix={arXiv},
      primaryClass={cs.CY},
      url={https://arxiv.org/abs/2407.12687}, 
}

@misc{wang2023mint,
      title={MINT: Evaluating LLMs in Multi-turn Interaction with Tools and Language Feedback},
      author={Xingyao Wang and Zihan Wang and Jiateng Liu and Yangyi Chen and Lifan Yuan and Hao Peng and Heng Ji},
      year={2023},
      eprint={2309.10691},
      archivePrefix={arXiv},
      primaryClass={cs.CL}
}

@article{yang2023intercode,
  title={Intercode: Standardizing and benchmarking interactive coding with execution feedback},
  author={Yang, John and Prabhakar, Akshara and Narasimhan, Karthik and Yao, Shunyu},
  journal={Advances in Neural Information Processing Systems},
  volume={36},
  pages={23826--23854},
  year={2023},
  url = {https://proceedings.neurips.cc/paper_files/paper/2023/hash/4b175d846fb008d540d233c188379ff9-Abstract-Datasets_and_Benchmarks.html}
}

@inproceedings{
    han2025convcodeworld,
    title={ConvCodeWorld: Benchmarking Conversational Code Generation in Reproducible Feedback Environments},
    author={Hojae Han and seung-won hwang and Rajhans Samdani and Yuxiong He},
    booktitle={The Thirteenth International Conference on Learning Representations},
    year={2025},
    url={https://openreview.net/forum?id=rpouyo09V0}
}

@article{li2020distinguishes,
  title={What distinguishes great software engineers?},
  author={Li, Paul Luo and Ko, Amy J and Begel, Andrew},
  journal={Empirical Software Engineering},
  volume={25},
  number={1},
  pages={322--352},
  year={2020},
  publisher={Springer}
}

@inproceedings{hewner2010game,
  title={What game developers look for in a new graduate: interviews and surveys at one game company},
  author={Hewner, Michael and Guzdial, Mark},
  booktitle={Proceedings of the 41st ACM technical symposium on Computer science education},
  pages={275--279},
  year={2010}
}

@inproceedings{begel2008pair,
  title={Pair programming: what's in it for me?},
  author={Begel, Andrew and Nagappan, Nachiappan},
  booktitle={Proceedings of the Second ACM-IEEE international symposium on Empirical software engineering and measurement},
  pages={120--128},
  year={2008}
}

@article{yang2024report,
  title={Report cards: Qualitative evaluation of language models using natural language summaries},
  author={Yang, Blair and Cui, Fuyang and Paster, Keiran and Ba, Jimmy and Vaezipoor, Pashootan and Pitis, Silviu and Zhang, Michael R},
  journal={arXiv preprint arXiv:2409.00844},
  year={2024}
}

@book{twidale2013over,
  title={Over-The-Shoulder Learning: supporting brief informal learning embedded in the work context},
  author={Twidale, Michael},
  year={2013},
  publisher={Graduate School of Library and Information Science}
}

@ARTICLE{green2024quality,
  author={Green, Collin and Jaspan, Ciera and Hodges, Maggie and Lin, Jessica},
  journal={IEEE Software}, 
  title={Developer Productivity for Humans, Part 7: Software Quality}, 
  year={2024},
  volume={41},
  number={1},
  pages={25-30},
  doi={10.1109/MS.2023.3324830}}
\end{document}